\documentclass[12pt]{article}

\usepackage{scicite}
\usepackage{amsmath}
\usepackage{amssymb}
\usepackage{graphicx}
\usepackage{bm}
\usepackage{upgreek}
\usepackage{times}
\usepackage{epstopdf}
\usepackage{color}
\usepackage{ulem}
\usepackage{endnotes}
\usepackage{eucal}
\usepackage[dvipsnames]{xcolor}

\newenvironment{sciabstract}{%
\begin{quote} \bf}
{\end{quote}}

\title{Sensing orbital hybridization of graphene-diamond interface with a single spin}

\author
{Yucheng Hao$^{1,2,\dag}$,
	Zhiping Yang$^{1,2,\dag}$,
	Zeyu Li$^{3,4,\dag}$,
	Xi Kong$^{5,\ast}$,
	Wenna Tang$^{5}$,\\
	Tianyu Xie$^{1,2}$,
	Shaoyi Xu$^{1,2}$,
	Xiangyu Ye$^{1,2}$,
	Pei Yu$^{1,2}$,
	Pengfei Wang$^{1,2,6}$,\\
	Ya Wang$^{1,2,6}$,
	Zhenhua Qiao$^{3,4}$,
	Libo Gao$^{5}$,
	Jian-Hua Jiang$^{7,8}$,\\
	Fazhan Shi$^{1,2,6,9\ast}$,
	Jiangfeng Du$^{1,2,6,10\ast}$\\
\\
\normalsize{$^{1}$ CAS Key Laboratory of Microscale Magnetic Resonance and School of Physical Sciences,}\\
\normalsize{University of Science and Technology of China, Hefei, 230026, China}\\
\normalsize{$^{2}$ CAS Center for Excellence in Quantum Information and Quantum Physics,}\\
\normalsize{University of Science and Technology of China, Hefei 230026, China}\\
\normalsize{$^{3}$ CAS Key Laboratory of Strongly-Coupled Quantum Matter Physics and Department of}\\
\normalsize{ Physics, University of Science and Technology of China, Hefei, 230026, China}\\
\normalsize{$^{4}$ The International Center for Quantum Design of Functional Materials,}\\
\normalsize{University of Science and Technology of China, Hefei, 230026, China}\\
\normalsize{$^{5}$ National Laboratory of Solid State Microstructures and Department of Physics,}\\
\normalsize{Nanjing University, Nanjing, 210093, China}\\
\normalsize{$^{6}$ Hefei National Laboratory, University of Science and Technology of China,}\\
\normalsize{Hefei, 230088, China}\\
\normalsize{$^{7}$ School of Physical Science and Technology \& Collaborative Innovation Center}\\
\normalsize{of Suzhou Nano Science and Technology, Soochow University, Suzhou, 215006, China}\\
\normalsize{$^{8}$ Suzhou Institute for Advanced Research, University of Science and Technology of China,} \\
\normalsize{Suzhou, 215123, China}\\
\normalsize{$^{9}$ School of Biomedical Engineering and Suzhou Institute for Advanced Research,}\\
\normalsize{University of Science and Technology of China, Suzhou 215123, China}\\
\normalsize{$^{10}$ School of Physics, Zhejiang University, Hangzhou 310027, China}\\
\\
\normalsize{$^\dag$ These authors contributed equally to this work.}\\
\normalsize{$^\ast$ Corresponding author. E-mail: kongxi@nju.edu.cn; fzshi@ustc.edu.cn; djf@ustc.edu.cn.}\\
}

\date{}

\begin{document} 


\baselineskip24pt


\maketitle


\begin{sciabstract}
Interfacial interactions are crucial in a variety of fields and can greatly affect the electric, magnetic, and chemical properties of materials. 
Among them, interface orbital hybridization plays a fundamental role in the properties of surface electrons such as dispersion, interaction, and ground states. 
Conventional measurements of electronic states at interfaces such as scanning tunneling microscopes are all based on electric interactions which, however, suffer from strong perturbation on these electrons. 
Here we unveil a new experimental detection of interface electrons based on the weak magnetic interactions between them and the nitrogen-vacancy (NV) center in diamond. 
With negligible perturbation on the interface electrons, their physical properties can be revealed by the NV spin coherence time. 
In our system, the interface interaction leads to significant decreases in both the density and coherence time of the electron spins at the diamond-graphene interface. 
Furthermore, together with electron spin resonance spectra and first-principle calculations, we can retrieve the effect of interface electron orbital hybridization.  
Our study opens a new pathway toward the microscopic probing of interfacial electronic states with weak magnetic interactions and provides a new avenue for future research on material interfaces.  
\end{sciabstract}

Interfaces of materials have attracted extensive research interest due to their exotic properties and play a prominent role in physics, chemistry and biology\cite{chandler_interfaces_2005,reyren_superconducting_2007,hellman_interface-induced_2017,li_interfacial_2019,barthelat_structure_2016}. 
However, the inherent complexity of interfaces, particularly the interaction between interfaces that determines electron correlation, spin-orbit coupling, orbital hybridization and charge redistribution\cite{mikkelsen_hybridization_2018,grisolia_hybridization-controlled_2016,li_approaching_2023}, is not well understood.
Two-dimensional (2D) materials\cite{novoselov_2d_2016,liu_2d_2019}---atomic or molecular interfaces widely used in science and applications---have been studied over the past decades with great interest. 
As one of the prototypes, graphene\cite{novoselov_electric_2004} has unsaturated bonds formed by $sp^2$ hybridization, which gives it excellent electrical properties but also makes it susceptible to substrates, contacts, and interfaces.
Understanding interfacial interactions and their effects is critical for improving the performance of graphene and other 2D material-based functional devices.
Yet, there is a lack of detection methods to directly study the hybridization of graphene interface at nanoscale under ambient conditions.
Scanning tunneling microscopy\cite{weiss_direct_2010}, transmission electron microscopy\cite{dou_atomic_2019} and atomic force microscopy\cite{zhang_real-space_2013} are powerful tools for probing intramolecular interactions, but they are incapable of studying hybridization through the graphene layer.
Angle-resolved photoemission spectroscopy\cite{zhou_substrate-induced_2007} has been used to experimentally probe the effects of hybridization on the energy band but is unable to be used in the nanoscale range.
In addition, due to the limitation of low sensitivity, most of the current experiments\cite{zhou_substrate-induced_2007,dou_atomic_2019} focus on graphene growing epitaxially on the substrate rather than being transferred from elsewhere, which is the most common usage scenario but difficult to detect owing to the weaker interactions.
The nitrogen-vacancy (NV) center\cite{maze_nanoscale_2008,balasubramanian_nanoscale_2008} in diamond is a well-known excellent sensor of its local electromagnetic environment.
Although the electron spins on the diamond surface cause decoherence of the shallow NV sensor\cite{romach_spectroscopy_2015,rosskopf_investigation_2014,dwyer_probing_2022}, these surface spins can be used as sensing reporters\cite{sushkov_magnetic_2014} and reactants that interact with substances on the diamond interface.
The NV-surface spins system provides an in situ and noninvasive probing technique that offers observations of electrons and their dynamics at the nanoscale.
The technique is well adapted for studying such interfacial interactions.

\begin{figure}
	\centering
	\includegraphics[width=0.9\columnwidth]{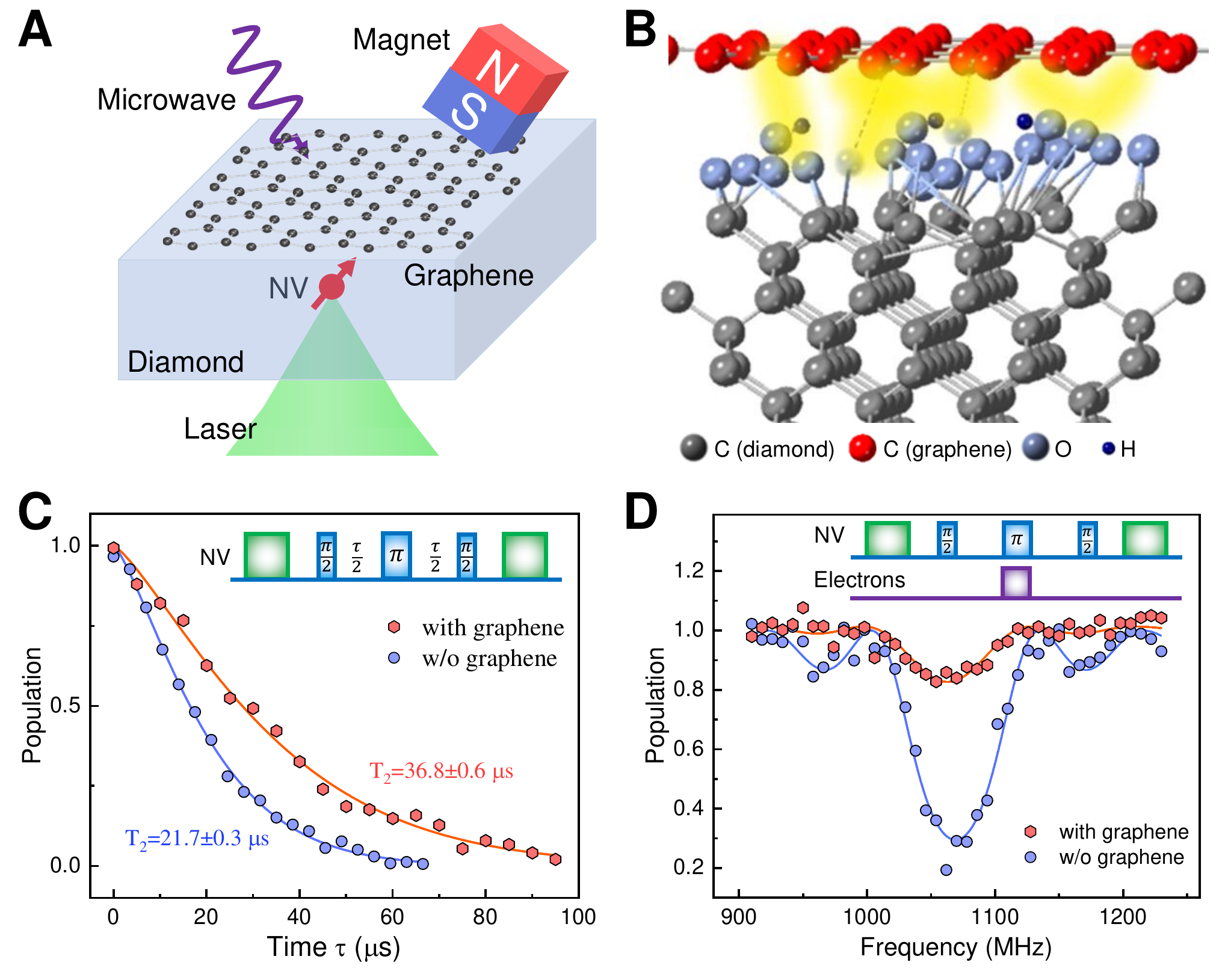}
	\caption{\textbf{Probing hybridization at the graphene-diamond interface.}
		\small
		{\textbf{(A)} Schematic illustration of the measurement. A single layer of graphene is transferred to the diamond surface. The shallow NV center is used as a sensor to detect the hybridization of the  graphene-diamond interface by probing the surface electrons.}  
		{\textbf{(B)} The unpaired electron spins on the diamond surface interact with the electrons of graphene when graphene is deposited on diamond.}   
		{\textbf{(C)} Hahn-echo measurement demonstrates an enhancement of the coherence time from 21.7 $\pm $ 0.3   $ \upmu\mathrm{s}$ to  36.8 $\pm $ 0.6 $ \upmu \mathrm{s}$. All of the experiments are performed at 382 Gauss, and the data are fitted by $e^{-(t/T_2)^p}$.}  
		{\textbf{(D)} Normal DEER spectra of the diamond sample in air and deposited with graphene. The pulse duration time is fixed while the frequency of the flip pulse (purple) of electrons is swept from 910 to 1230 MHz. Solid lines are fitted to the analytical solutions under semiclassical approximation.}  
	}\label{fig:1}
\end{figure}

As a highly sensitive magnetic sensor\cite{xie_beating_nodate}, the NV center is used to measure the electron spins on the diamond surface. 
Here we report that the density and relaxation time of these electron spins at the diamond surface can be modulated by introducing the graphene-diamond interface, which is discovered due to the orbital hybridization at the graphene-diamond interface.
The schematic of the experimental principle is illustrated in Fig. \ref{fig:1}A.
The diamond is a 50 $\upmu$m $\left\langle 100\right\rangle $-oriented film, including a 10 $\upmu \mathrm{m}$ $^{12} \mathrm{C}$-enriched layer.
NV centers located at a few to 20 nm below the diamond surface are created by ${ }^{15} \mathrm{~N}^{+}$ low dosage ion implantation and subsequent annealing.
Monolayer graphene was directly deposited on the diamond after a series of sample-free measurements.  
The distance between graphene and diamond surface is on the nanometer scale, which can be verified by observing the decay of NV fluorescence intensity (fig. S1) due to F{\"o}rster resonance energy transfer\cite{tisler_single_2013}.
Such that the electrons at both interfaces may undergo hybridization interaction (Fig. \ref{fig:1}B).
By comparing the surface electron signals detected by the NV center before and after the deposition of graphene, it is possible to determine the hybridization property of the graphene-diamond interface.

In our experiments, the magnetic noise of electron spins on the diamond surface affected the decoherence process of the shallow NV center.
The coherence time of NV centers was measured by the Hahn-echo experiment.
Fig.\ref{fig:1}C shows the coherence time $T_2$ of NV1 without and with graphene at an external magnetic field $B_0$ = 382 Gauss along the symmetry axis.
The coherence time $T_2$ was   21.7 $\pm $ 0.3 $ \upmu\mathrm{s}$ when the diamond surface was directly exposed to air.
As the depth of NV1 was expected to be around 10 nm, the decoherence was mainly originated from the electrons on the surface.
With the graphene deposited, $T_2$ increased to  36.8 $\pm $ 0.6 $ \upmu \mathrm{s}$, which indicated that the surface noise was suppressed.
This is also verified by comparing the electron spin resonance spectra of surface electrons, which can be measured by double electron-electron resonance (DEER) experiments\cite{grotz_sensing_2011}.
Each spectrum in Fig. \ref{fig:1}D displays a clear dip at the resonance frequency of $g = 2$ dark electrons without graphene, corresponding to the electron spins.
Remarkably, the amplitude of the electron spin resonant dip was greatly reduced when covered with graphene.
This indicates that the graphene reduces the detected surface electron signal originating from the interface.
The same measurements have been performed on different NV centers, and most of them found an increase in the coherence time accompanied by a decrease in the amplitude of the electron spin spectra\cite{som}.

\begin{figure}
	\centering
	\includegraphics[width=1\columnwidth]{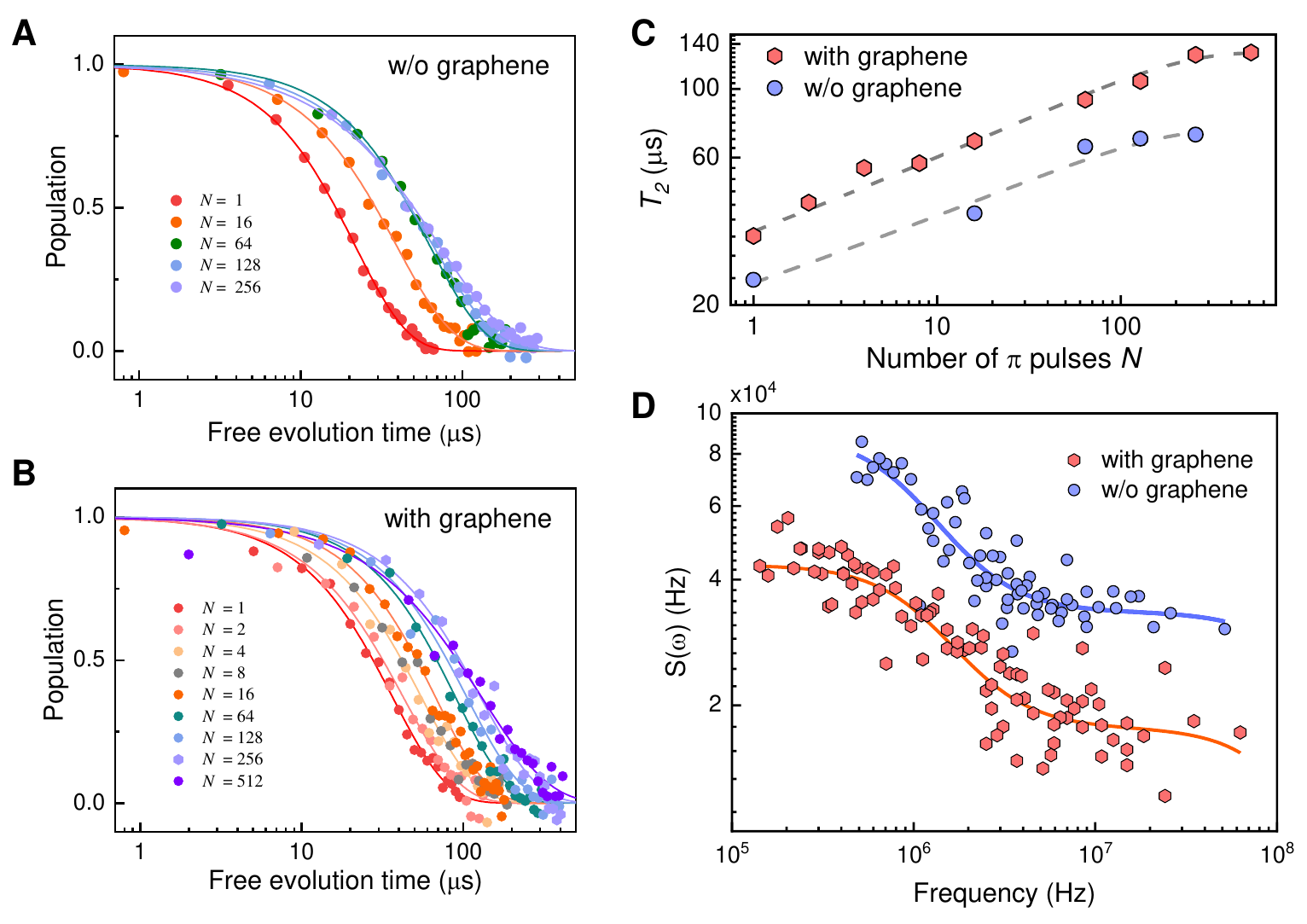}
	\caption{\textbf {The decoherence curves and noise spectra for NV1 with and without graphene deposited.}
		\small
		{\textbf{(A)} Coherence versus time for different numbers of $\uppi$ pulses without graphene.} 
		{\textbf{(B)} Coherence versus time for different numbers of $\uppi$ pulses with graphene deposited.} 
		{\textbf{(C)} Coherence time as a function of the number of $\uppi$ pulses of the dynamical decoupling sequence with and without graphene.} 
		{\textbf{(D)} Noise spectrum derived from decoherence  without (A) and with (B) graphene. The solid lines are fitted by the double Lorentzian curve. }
	}\label{fig:2}
\end{figure}

To gain further insight into the mechanism of interface modification of diamond and graphene, the decoherence behavior and noise spectrum of the NV center were investigated.
the periodic dynamical decoupling Carr-Purcell-Meiboom-Gill (CPMG) pulse sequences with varying number of $\pi$ pulses were applied until the coherence time was saturated.
The measured coherence as a function of time for NV1 with and without graphene are shown in Fig. \ref{fig:2}A and Fig. \ref{fig:2}B, respectively.
Then the data were fitted with $e^{-(t/T_2)^p}$ to extract the coherence time, from which we extracted the coherence time as a function of the numbers of $\pi$ pulses, as shown in Fig. \ref{fig:2}C.
It can be seen that the coherence time with graphene deposited is always longer than that without graphene for various orders of the sequences.
Then the noise spectra were obtained by spectral decomposition\cite{bar-gill_suppression_2012}, where the coherence data were deconvolving with the filter function of each pulse sequence\cite{som}.
As shown in Fig. \ref{fig:2}D, the two noise spectra exhibited overall similar behaviors, but the noise intensity decreased significantly with graphene deposited, exhibiting the ability of graphene to suppress noise in the resolved frequency range.
We fitted the noise spectrum with the double Lorentzian function and found that the higher frequency noise which saturates the coherence time and may result from the surface modified phonons\cite{romach_spectroscopy_2015} was also suppressed.

\begin{figure}
	\centering
	\includegraphics[width=1\columnwidth]{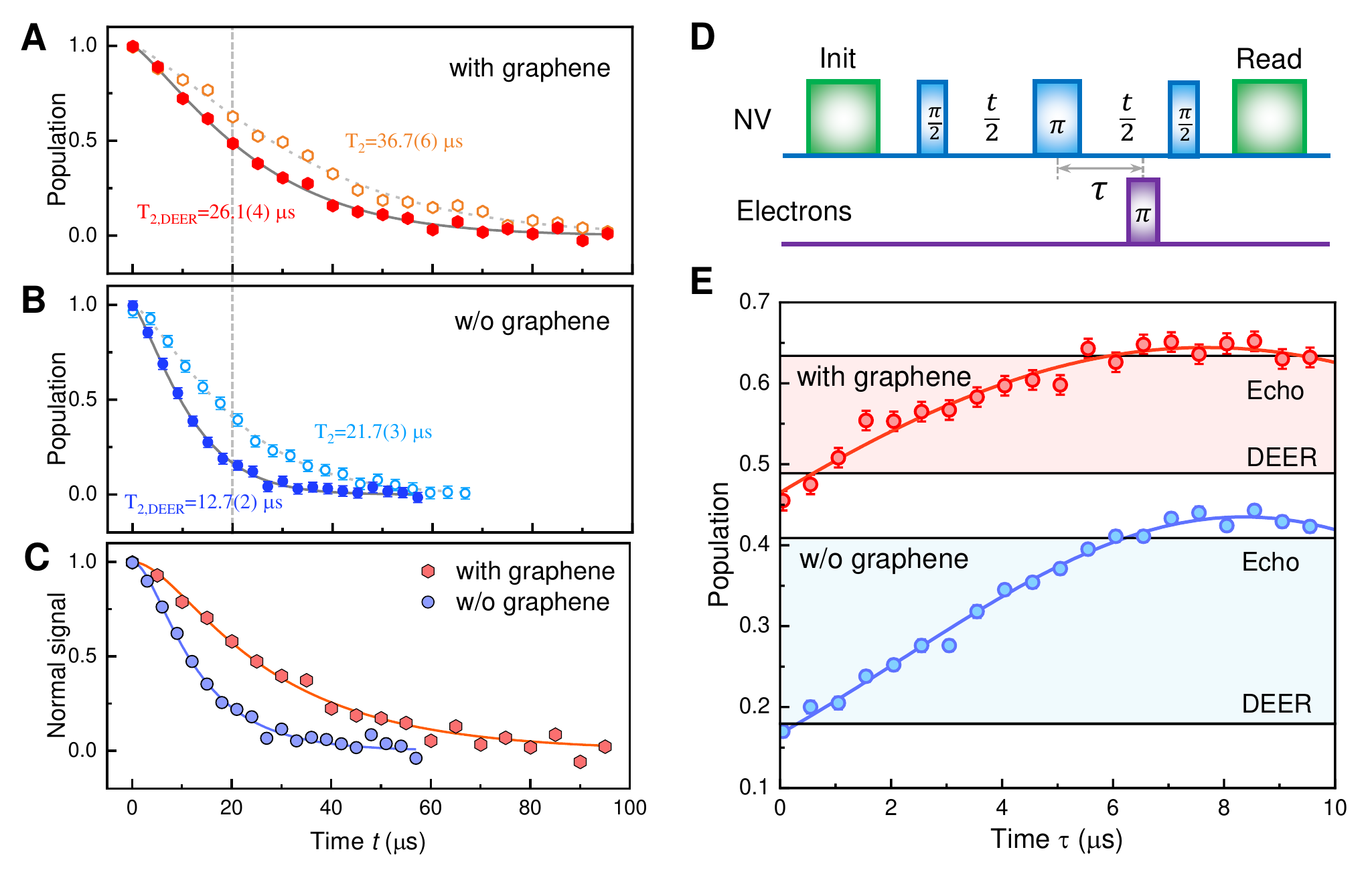}
	\caption{\textbf{Quantitative measurement of the density and longitudinal relaxation time of the surface electrons by NV 1.}
		\small
		{\textbf{(A)} The DEER and $T_2$ decoherence curves without graphene.} 
		{\textbf{(B)} The DEER and $T_2$ decoherence curves with graphene deposited.} 
		{\textbf{(C)} The DEER decoherence curves are normalized with the noise floor extracted from the noise spectrum to exclude decoherence due to the faster noise, which shows that the coupling strength between the NV center and the surface spin bath was decreased from $90.9\pm5 $ kHz to $41\pm 0.4$ kHz.}  
		{\textbf{(D)} Schematic of the DEER-echo sequence to probe the density and relaxation time of electron spins.}  
		{\textbf{(E)} The DEER-echo curves without (blue) and with (red) graphene deposited. The fitting results show that the electron spin density decreased from $0.0025\pm 0.0002$ nm$^{-2}$ to $0.0011\pm 0.0002$ nm$^{-2}$, while the relaxation time turned from $21\pm3$ $\upmu \mathrm{s}$ to $12\pm 3$ $\upmu \mathrm{s}$.}
	}\label{fig:3}
\end{figure}

The increase in coherence time and the decrease in noise level indicated that the properties of the interfacial electron spins change significantly. 
The electron spins were further investigated before and after graphene deposition. 
The DEER and Hahn-echo experiments were measured as shown in Fig. \ref{fig:3}A and Fig. \ref{fig:3}B. 
It is clear that the difference between the DEER and Hahn-echo decoherence curves decreased as the coherence time increased with graphene deposited.
As mentioned above, the surface-modified phonons give rise to the higher frequency noise and lead to a noise floor in the resolved frequency\cite{romach_spectroscopy_2015} (Fig. \ref{fig:2}D).
We then excluded the effect of the phonons by normalizing the deer decoherence curves with the faster noise extracted from the noise spectra, as shown in Fig. \ref{fig:3}C.
The curves were fitted by $	c(t)=e^{-\gamma_e^{2}B_{rms}^{2} \tau_{c}^{2}\left(e^{-t / \tau_{c}}+t / \tau_{c}-1\right)}$, where $	B_{rms}$ is the fluctuating magnetic field induced by electron spins and $\tau_c$ is the electron relaxation time.
However, the DEER method could not extracted  $	B_{rms} $ and $\tau_c$  exactly because it is challenging to remove the decoherence irrelevant to the surface spin bath (fig. S2).
The fitting results show that the product of these two factors $\gamma_e^{2} B_{rms}^2 \tau_c$, called coupling strength, decreased from 90.9 $\pm $ 5 kHz to 41 $\pm$ 0.4 kHz.

To further reveal the essence of the interaction mechanism, it is necessary to distinguish the contributions between the density and relaxation time of the interface electrons.
The correlation spectroscopy sequence\cite{sushkov_magnetic_2014} based on DEER can be used to probe the relaxation time of the surface electrons, but the sensitivity is not sufficient to acquire a comparatively accurate result within the effect time, especially when the coverage of graphene lowers the readout efficiency and shortens the relaxation time of electrons.
An experimental sequence DEER-echo\cite{mamin_detecting_2012} is shown in Fig. \ref{fig:3}D.
The phase accumulation time $t$  of the spin-echo is fixed and the interval time $\tau$ between the two microwave $\pi$ pulses is changed, so that decoherence due to anything other than spin noise is a constant with varying $\tau$ and can be easily excluded, and the spin density and relaxation time can be clearly extracted by fitting to the theoretical model Eq.S16 in \cite{som}.
As shown in Fig. \ref{fig:3}E, the coherence is consistent with the DEER experiment at $\tau$ = 0  $\upmu \mathrm{s}$ and with the Hahn-echo measurement at $\tau$ = 10  $\upmu \mathrm{s}$, in which the sequence degenerates to DEER or Hahn-echo in such conditions, as shown by the dash line in Fig. \ref{fig:3}A, B.
It is worth noting that a distinct peak (coherence exceeds the Echo line) appears during the period in each measurement, which shows the susceptibility of the curve shape to the relaxation time $\tau_{c}$.
The extracted results show that the electron spin density decreased from $0.0025\pm 0.0002$ nm$^{-2}$ to $0.0011\pm 0.0002$ nm$^{-2}$, while the relaxation time turned from $21\pm3$ $\upmu \mathrm{s}$ to $12\pm 3$ $\upmu \mathrm{s}$.
Besides, we haved observed that the decrease of  electron spin density and relaxation time on the most of NV centers we measuremed, see Table S1 and fig. S8 in \cite{som}.


\begin{figure}
	\centering
	\includegraphics[width=1\columnwidth]{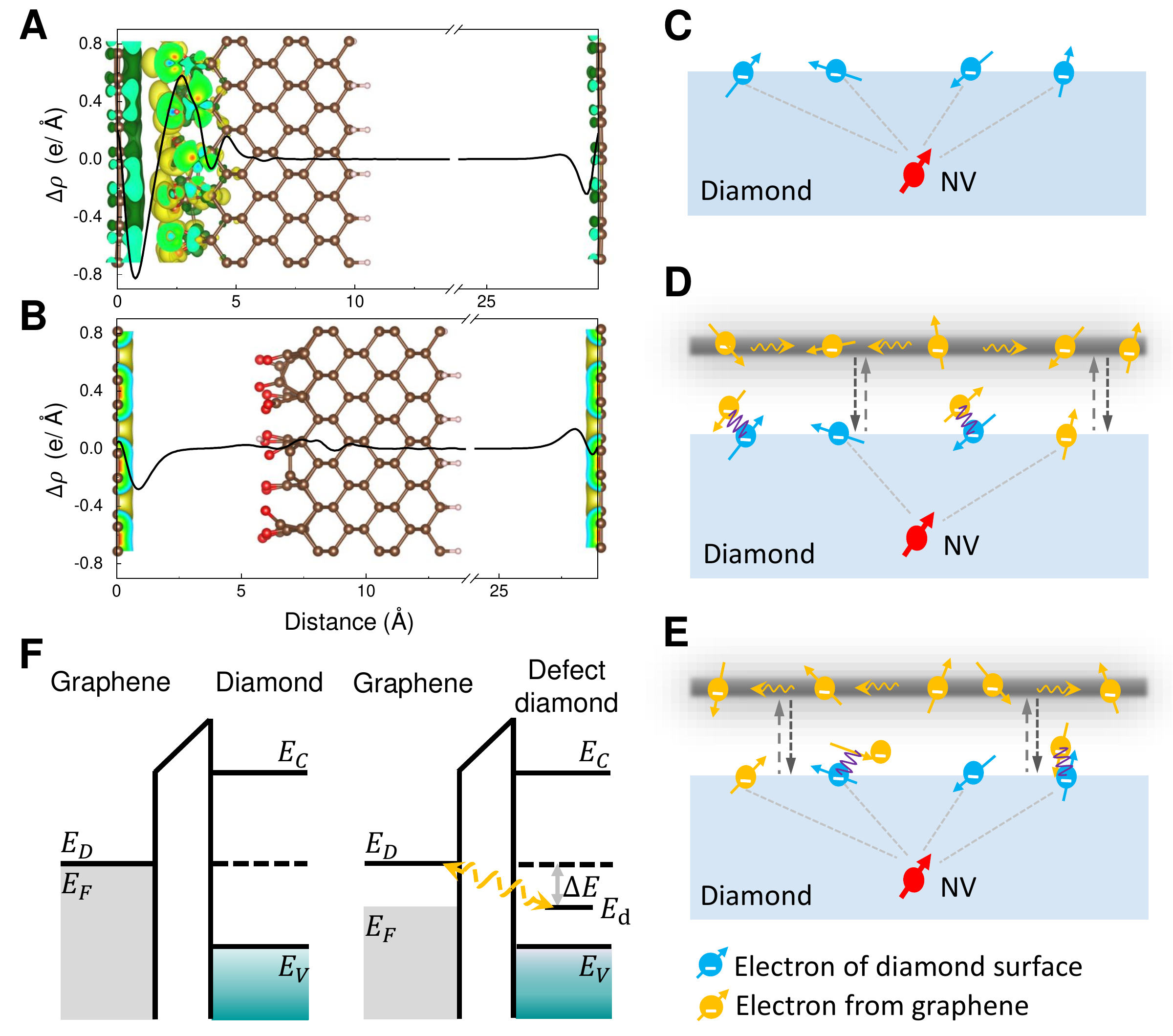}
	\caption{\textbf{Theoretical analysis of the graphene and diamond interface.}		
		\small
		{\textbf{(A, B)} The visualized charge density difference and planar-average charge density along the z-direction with a distance of graphene diamond d = 2.3 \AA  \ in (A) and d = 6 \AA  \ in (B). The isosurface value is 0.00079 $e/Bohr^3$. } 
		{\textbf{(C, D, E)} Schematic showing the interaction of interface electrons before (C) and after (D, E) graphene deposition. The unpaired electron spins on the diamond contribute to the decoherence of NV center.} 
		{\textbf{(F)} Left: Energy diagram of graphene and diamond when they are far away or not interacting. Right:  Energy diagram of graphene and diamond with surface defects shows a refined Fermi level and allows charge flows from graphene to diamond.} 
	}\label{fig:4}
\end{figure}

To elucidate the effects of graphene on diamond, the charge distribution and orbital projected band structure of the graphene-diamond heterostructure are analyzed by ab-initio calculations. 
A graphene monolayer is deposited on the surface of $ <$100$> $-oriented diamond with a distance of approximately 2.3 \AA  \ after full relaxation, accompanied by residual bonds such as C-, O- and OH- terminals at the diamond surface.
The charge redistribution at the interface is illustrated by the charge density difference, which is defined as $\delta \rho = \rho_{\rm dia/gra}-\rho_{\rm dia}-\rho_{\rm gra}$, where $\rho_{\rm dia/gra}$, $\rho_{\rm dia}$ and $\rho_{\rm gra}$ are the charge densities of the diamond/graphene vdW heterostructure, diamond, and graphene, respectively.
Fig. \ref{fig:4}A shows the charge density map of the system, with the black solid line corresponding to the planar-average charge density along the z-direction.
It is obvious that the charge density redistribution and even overlap are formed in the vdW gap.
The charge accumulation regions are close to the side of diamond, demonstrating that electrons can be transferred from graphene to diamond  and allow interactions to occur.
In contrast, the charge density map with a distance of 6 \AA  \ of the interface is shown in Fig. \ref{fig:4}B.
The charge density map shows that the electrons just accumulate at the surface of graphene without redistribution once the space is large enough.
We also calculated the charge transfer of the graphene-diamond interface with different initial distances by using Bader charge analysis \cite{tang_grid-based_2009}(fig. S13).
The results showed that the transferred electrons monotonically decrease with increasing the distance, which could explain the difference in density variations for each NV.
In addition, the projected band structure results suggest that the presence of the diamond substrate has a negligible effect on the energy structure of graphene.
For the structures with these defects, the band structure that only comprises p orbitals of diamond exhibits typical features of graphene gapped Dirac cones (fig. S14).

Thus the changes can be explained when we consider the orbital hybridization of electrons between the graphene-diamond interface.
When the diamond was not covered with graphene, the unpaired electrons on the diamond surface relaxed due to  the surrounding noise and were detected by the NV center (Fig. \ref{fig:4}C).
When the graphene was tightly adhered to the diamond, their interactions redefined the position of the Fermi level and the electrons in the graphene tended to the diamond surface and hybridize with the unpaired electrons of the diamond (Fig. \ref{fig:4}D,E).
Due to our room temperature conditions, the electrons kept transitioning back and forth between graphene and diamond surface bound states, leading to a decrease in the electron relaxation time, as shown in the energy diagrams of graphene and diamond with or without surface defects in Fig. \ref{fig:4}F.
Some highly hybridized graphene-diamond surface electrons paired and formed covalent bonds of spin singlet states, thus reducing the density of electron spins that can be detected by NV.
In addition, it is well known that the mobility of electrons in graphene is fast with a relaxation time estimated to be the ps level\cite{du_approaching_2008,pogna_electrically_2022}, which is not detectable and thus does not contribute to the decoherence of the NV in our experiments.

In summary, we experimentally observed the graphene-diamond interface interactions at the nanoscale by NV sensors. 
We found that the coherent properties of NV centers were significantly enhanced and the detected surface electron signal was reduced after graphene was deposited on diamond.
This is beneficial for quantum sensing with increased sensitivity, and since graphene is only a single atomic layer thick, it does not affect the detection of external signals.
Further experimental results and analysis indicated that the hybridization occurred at the interface between diamond and graphene, leading to a decrease in both the density and relaxation time of surface electron spins.
It can be inferred that the microscopic mechanism can interpret the changes in the macroscopic properties, such as contact potential and conductivity\cite{datta_crystallographic_2008,shi_effective_2009}.
Our study demonstrates the possibility of improving the performance of graphene-derived devices, which is important for the study of graphene-based devices and has a wide range of semiconductor applications. 
In addition, the reduction in the high frequency shown in the reconstructed noise spectrum deserves further investigation, and the suppression of interface noise may be enhanced by a more effective contact between the two surfaces.
Although interfacial interactions have been widely studied and exploited, our new approach allows for nondestructive measurements of weak phenomena at the nanoscale, which is difficult to realize by other existing methods.
Future work can be used to explore the kinetic behavior of interfacial materials and play a role in the field of interfacial catalysis\cite{li_interfacial_2019}, superconductivity\cite{kamlapure_tuning_2022} and topological surface states\cite{hsu_hybridization-induced_2017}.

\bibliography{graphene_ref.bib}

\bibliographystyle{Science}

\section*{Acknowledgments}
\textbf{Funding:} This work is supported by 
the National Natural Science Foundation of China (grant nos. T2125011, 81788101, 12125504, 12274396), 
the National Key $R \& D$ Program of China (grant nos. 2018YFA0306600), 
the CAS (grant nos. XDC07000000, GJJSTD20200001, QYZDY-SSW-SLH004, Y201984, YSBR-068), 
Innovation Program for Quantum Science and Technology (Grant No. 2021ZD0302200, 2021ZD0303204), 
the Anhui Initiative in Quantum Information Technologies (grant no. AHY050000), 
Hefei Comprehensive National Science Center, 
and the Fundamental Research Funds for the Central Universities.
This work was partially carried out at the USTC Center for Micro and Nanoscale Research and Fabrication.

\textbf{Author contributions:}  
	J.D. and F.S. supervised the project and proposed the idea. 
J.D., F.S. and X.K. designed the experiments. 
Z.Y., Y.H. and X.K. constructed the experimental setup.
Y.H. and Z.Y. performed the experiments, data acquisition and calculation.
X.Y., P.Y. and Y.W. prepared the diamond sample. 
W.T. and L.G. prepared the graphene sample. 
Z.L. and Z.Q performed first-principles calculations.
J.J. inspired us and discussed on the theoretical analysis. 
Y.H., Z.Y., X.K., T.X., J.J., and F.S. performed the data analysis and interpretation.
Z.Y., Y.H., X.K., Z.L., J.J., and F.S. wrote the manuscript. 
All authors discussed the results and commented on the manuscript.
\textbf{Competing interests:} The authors declare no competing interests. 
\textbf{Data and materials availability:} All data are available in the manuscript or the supplementary materials.
\section*{Supplementary materials}
Materials and Methods\\
Supplementary Text\\
Figs. S1 to S14\\
Table S1 \\
References \textit{(35-45)}


\end{document}